\documentclass[a4paper, oneside, twocolumn, notitlepage, 10pt]{extarticle_ecoc}
\usepackage{ecoc}
\usepackage{soul}
\usepackage{comment}

\renewcommand\footnotemark{}

\begin{document}
\selectlanguage{english}    


\title{Maximum Achievable Burst Size in All-Optical Satellite Networks}


\author{
    Thomas Roethig\textsuperscript{(1)}, Sleman Mouammar\textsuperscript{(1)},
    Ítalo Brasileiro\textsuperscript{(1)},
    Soheil Hosseini\textsuperscript{(1)},Admela Jukan\textsuperscript{(1)}
}

\maketitle                  


\begin{strip}
    \begin{author_descr}

        \textsuperscript{(1)} Technische Universit\"at Braunschweig, Germany,
        \textcolor{blue}{\uline{\{thomas.roethig, sleman.mouammar1\}@tu-bs.de}}
    \end{author_descr}
\end{strip}




\begin{strip}
    \begin{ecoc_abstract} 
We analyze the maximum burst size achievable in all-optical satellite networks across different constellations. With a 100 Gbps uplink capacity, a WDM-based optical burst switching network supports burst sizes of up to 500 MB in high-altitude LEO constellations and 600 MB in low-altitude LEO constellations. $\copyright$ 2026 The Author(s)
    \end{ecoc_abstract}
\end{strip}


\section{Introduction}
\begin{figure*}[b]
    \centering
    \includegraphics[width=0.95\linewidth]{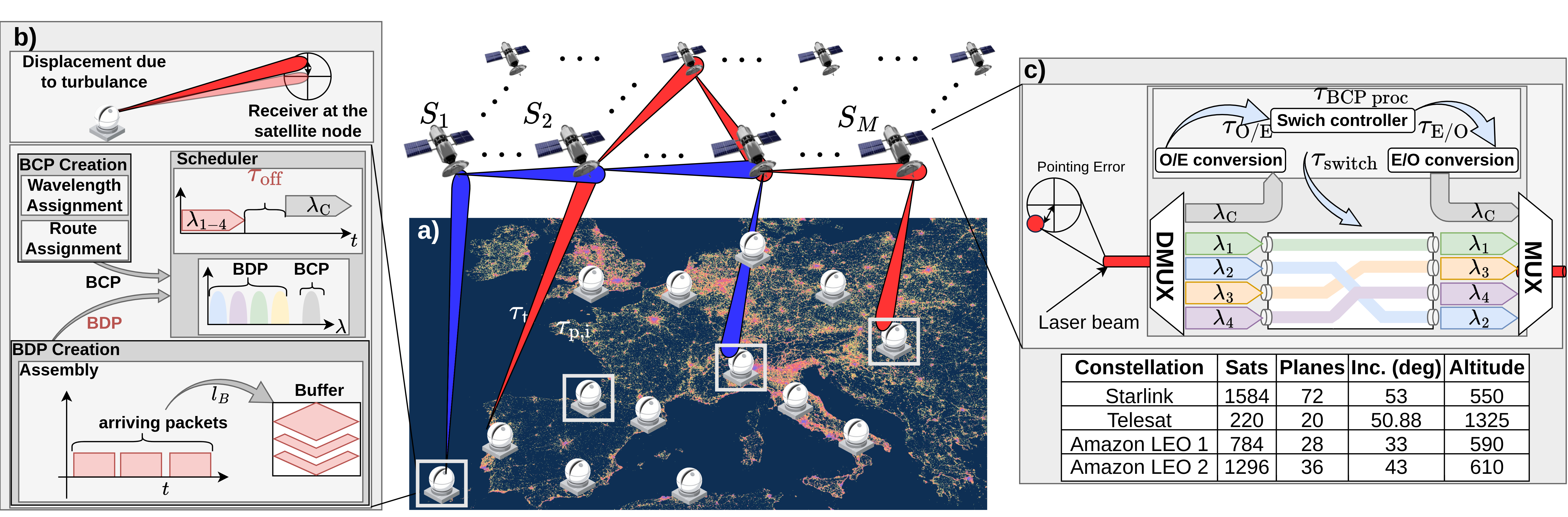}
    \caption{a): Reference scenario, consisting of $N$ OGSs, and M LEO satellites. b) zoom in on the burst assembly at the OGS; c) zoom in on the all-optical satellite nodes, and the used constellations.}
    \label{fig:arc}
\end{figure*}
Today's satellite networks are rapidly adopting optical solutions for uplink, downlink, and laser inter-satellite link (LISL) connections~\cite{Cardakli:26,Armengol:23}. To fully utilize satellite link capacities of hundreds of Gb/s, all-optical switching paradigms based on wavelength-division multiplexing (WDM), similar to those used in terrestrial networks, can be considered \cite{Karafolas02}. To this end, Optical Burst Switching (OBS) was revisited as a possible all-optical switching paradigm for satellite networks \cite{Zhai:21}, offering the potential for better link utilization than circuit switching. In OBS, a burst data packet (BDP) is assembled at the optical ground station (OGS) and transmitted over a pre-reserved optical end-to-end (E2E) path established by the burst control packet (BCP).

While OBS literature in terrestrial networks is rich, OBS performance across varying constellations is not yet well understood. Previous studies have validated OBS for satellite payloads as in \cite{Zhai:21}, where the system was presented in a tabletop experiment employing an arrayed waveguide grating router (AWGR) for transparent optical switching, along with 10-Gbps tunable lasers and wavelength-conversion capabilities. It was shown that the implementation is constrained by the minimum switching interval of 11.9 $\mu$s, which limits reconfiguration speed. The work in \cite{zhao2022distributed} investigated the impact of burst size in a GEO satellite network restricted to path lengths of 1 to 3 hops. Simulations considered single channel rates of 10, 40, and 100 Gbps and burst durations of 100, 200, and 500 $\mu$s. It was shown that OBS performance is superior to the traditional optical circuit switching (OCS). 

Motivated by OBS’s advantages over conventional optical switching, this paper analyzes the maximum burst size that can be assembled at the OGS in bufferless multi-channel OBS satellite networks under realistic 30-60 ms E2E latency constraints \cite{dano2026ookla}. We assess through simulation various burst sizes in real-world constellations such as Starlink, Telesat, and Amazon LEO. The results show that burst sizes of 300–500~MB meet the target latency requirement, while having little impact on burst loss for a given load. In contrast, E2E latency increases linearly with burst size, indicating that higher-altitude LEO constellations require shorter bursts to meet the latency requirements. To the best of our knowledge, this is the first paper to analyze feasible burst sizes in all-optical LEO constellations leveraging the OBS paradigm.


\section{Scenario Description}
Fig.~\ref{fig:arc}a) illustrates the reference scenario where the integration of all-optical LEO satellite constellations is expected to mitigate latency between two distant points in terrestrial networks. The reference scenario includes multiple OGS access points in high-demand areas. Within the scope of our evaluation, we consider a total of four OGSs as traffic hotspots. Their placement is based on real operational sites, prior work in the literature, and traffic hotspot regions identified from available data. Accordingly, two real-world operational OGSs are located in Maspalomas (27.76$^\circ$,-15.59$^\circ$) and Athens (37.98$^\circ$,23.72$^\circ$), while two additional OGSs are placed in Milan (45.4642$^\circ$, 9.1900$^\circ$) and Bilbao (43.2630$^\circ$, -2.9350$^\circ$), following \cite{Farley:24}. To identify these hotspots, we use the Ookla and Human Settlement Proxy datasets, combining latency measurements with urban development indicators to capture current and future high-demand areas. 

We analyze four different LEO constellations, namely one high altitude (Telesat), which ensures longer visibility windows for up- and downlinks but at the cost of longer propagation, and 3 lower altitude ones (Starlink, Amazon LEO 1 \& 2) with variable sizes, to assess the impact of constellation size and altitude (table in Fig.\ref{fig:arc} c)).

In the OGS, traffic to be sent over the satellite network is assembled into bursts (Fig.~\ref{fig:arc}b)). In our approach, burst assembly  followed a burst-size threshold of $l_B$ bytes. Once a burst is formed, the source OGS generates the corresponding BCP, which carries the route and assigned wavelength information. Routing is performed using the shortest-path algorithm, while wavelength assignment is handled by the \textit{Horizon} algorithm \cite{teng2005detailed}, which follows a scheduling policy to select a free wavelength with the earliest \textit{time horizon}, defined as the earliest time at which the wavelength becomes available after previously scheduled bursts. A burst is accepted for transmission only if its arrival time exceeds this horizon; once scheduled, the \textit{time horizon} is updated to the burst reservation's end time. The scheduler ensures that the BCP is transmitted ahead of the corresponding BDP by an offset time $\tau_{\mathrm{off}}$. This offset must exceed the satellite processing time and the transmission delay $\tau_{\mathrm{tr}}$, as illustrated in Fig.\ref{fig:arc}b). The BCP is sent over a dedicated wavelength $\lambda_C$, while the BDP is sent over one of four wavelengths $\lambda_1-\lambda_4$.

Once the scheduled BCP reaches the first satellite, as illustrated in Fig.~\ref{fig:arc}c), it undergoes O/E conversion $\tau_{\mathrm{O/E}}$, electrical processing $\tau_{\mathrm{BCP_{\mathrm{proc}}}}$, and E/O conversion $\tau_{\mathrm{E/O}}$. Containing the wavelength and route information, the BCP configures onboard switches so the BDP can be transparently switched after the offset time, enabling high-throughput, bufferless satellite data switching.

\section{Simulation Setup}

Event-driven simulations are implemented using OMNeT++ v6.3.0. The OS3~\cite{seregi2015os3} library in OMNeT++ is used to generate satellite constellations (from table in Fig. \ref{fig:arc} c)), including their propagation models derived from the Three Line Element (TLE) files obtained via~\cite{TLEGen}.  The OBS~\cite{izal2013obsmodules} library is adapted and integrated with OS3 to meet the requirements previously defined, such as multi-channel WDM links, an optical channel estimator, wavelength assignment, burst assembly, scheduler, and routing, as well as Grid+ LISLs. Each satellite node has four ports for neighboring satellites in a Grid+ topology, with average hop counts of 1.5, 2.5, 2.3, and 2.9 for Telesat, Amazon LEO 1, Amazon LEO 2, and Starlink, respectively, plus one additional port for OGS connectivity.

To characterize the physical-layer behavior of the allocated links, WDM channel links are simulated using the irradiance displacement model for LISLs from \cite{Farid,Shang:25}. The up- and downlinks are simulated by accounting for atmospheric turbulence $L_{atm}$, pointing error $L_{p}$, free-space path loss $L_{FSO}$, and the number of wavelengths $N_{ch}$ \cite{Spirito:25}. Accordingly, the received power is given by $P_R = \frac{P_T}{N_{ch}} , G_T \eta_T G_R \eta_R L_{FSO} L_{atm} L_{p}$, where $G_T$ and $G_R$ denote the transmitter and receiver gains, and $\eta_T$ and $\eta_R$ represent the optical efficiencies \cite{Spirito:25}. For LISLs, the power budget is defined analogously, but without the atmospheric loss term $L_{atm}$. The resulting signal-to-noise ratio is defined as $SNR = \frac{P_R}{N_0}$. Based on this, the channel capacity is given by $C = B \log_2(\mathrm{SNR}_k + 1), {k \in \mathrm{LISL,~uplink,~downlink}}$, where $\mathrm{SNR}_k$ is determined from the corresponding power budget.

To determine the maximum burst size, in each simulation, a fixed burst size is used, and varied between 1 MB and 1 GB across runs. Burst arrivals were modeled according to an Erlang distribution with parameters $k = 1$ and rate $\lambda_E$. To ensure a comparable load across different burst sizes, $\lambda_E$ was calculated from the bottleneck channel capacity and the transmission delay of the respective burst size. Consequently, the burst arrival rate is inversely proportional to the burst holding time, i.e., its transmission duration, such that smaller bursts lead to shorter inter-arrival times. Table~\ref{tab:table1} summarizes the remaining parameters.

Performance evaluation is then implemented based on the following two metrics: 
\par {\emph{E2E latency $L$:} The metric $L$ is defined as the sum of three components: i)  propagation delay $\tau_{\mathrm{p,i}}$ across all hops $i$ along the routing path $p$, ii)  satellite switching, processing, and O/E/O delays $\tau_{\mathrm{Sat,l}}$ at each hop $l$, and iii) transmission delay $\tau_{\mathrm{tr}}$ at the bottleneck link, i.e., 
$L= \sum_i^{\mathrm{Hop}} \tau_{\mathrm{p,i}} + \frac{\mathrm{Burst~Size}}{\mathrm{min}(C)} + \sum_l^{\mathrm{Hop}} \tau_{\mathrm{Sat,l}}$.

\par {\emph{Burst loss rate:}} This metric is defined as the ratio between the number of bursts not delivered and the total number of bursts sent, i.e.,
$\frac{\text{Bursts Lost}}{\text{Bursts Sent}} * 100$.
It occurs due to resource unavailability or line-of-sight disruptions in the up- and downlink.

\begin{table}[t]
    \centering
    \caption{Simulation parameters} \label{tab:table1}\scriptsize
    \begin{tabular}{|c|c|}
        \hline  \textbf{Parameter} & \textbf{Value} \\
        \hline Wavelength & 1550 nm \\
        \hline Rate ($\lambda$) & 0.026-2.6\\
        \hline Burst sizes & 1MB - 1GB \\
        \hline OGS aperture size& 0.8m \cite{Farley:24}\\
        \hline OGS gain & 125 Dbi \cite{Cantore:2024}\\
        \hline Satellite power & 0.5 W \cite{Shang:25} \\
        \hline Visibility & 60 km \cite{Spirito:25}\\
        \hline Nr. of WDM wavelengths & 5 per port \\
        \hline Nr. of ports per node & 4 (ISL) + 1 (OGS) bidirectional ports \\
        \hline Guard time & 100 ns \\
        \hline Switching speed & 5~ns \\
        \hline Processing + O/E/O time & (8 + 2) ns\\
        \hline Minimum elevation angle & $30^\circ$\\        
        \hline
    \end{tabular}
\end{table}

\section{Performance Evaluation}

\begin{figure}[ht]
    \centering
    \includegraphics[width=0.95\linewidth]{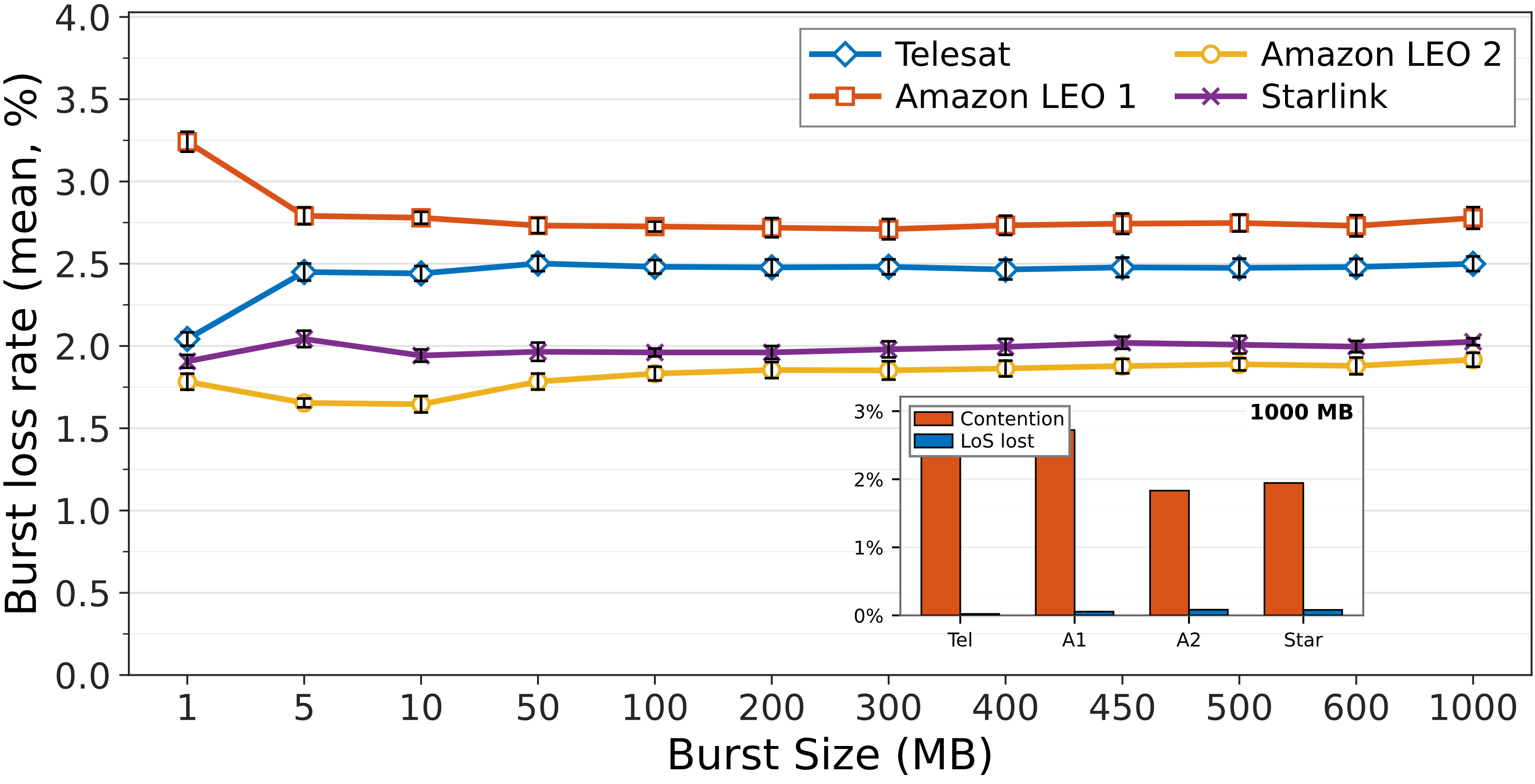}
    \caption{Burst loss and decomposition for different burst sizes and constellations.}
    \label{fig:blocking}
\end{figure}

Based on the E2E latency and burst loss rate,  the maximum feasible burst size is defined such that the E2E latency remains between 30 and 60 ms, over various constellations \cite{dano2026ookla}. Fig.~\ref{fig:blocking} presents the burst loss rate as a function of burst size across multiple LEO constellations. The burst loss rate is nearly constant when varying the burst size for each constellation. However, significant variations in loss are observed across different topologies, due to constellation-related parameters such as orbital altitude and satellite count. Amazon LEO 1, with a reduced satellite count and altitude, has reduced coverage, which extends the exposure of OGSs to the same shortest path for a longer period. This long-lasting route intensifies resource competition, increasing contention. In contrast, a denser satellite constellation such as Amazon LEO 2 enables more frequent updates of the selected path between OGSs, improving traffic distribution. Consequently, Amazon LEO 2 with 5MB (1.65\% loss) or 10MB (1.65\% loss) burst sizes achieves the lowest burst loss rate, representing a 49.07\% reduction relative to the worst-case scenario (Amazon LEO 1 with 1 MB bursts and 3.24\% loss). 

The Telesat constellation further demonstrates the coverage advantages of higher orbital altitudes. Despite operating with less than one-third the satellites of Amazon LEO 1, Telesat achieves an average burst loss reduction of 12.54\% (2.04\% for 1~MB) compared to Amazon LEO 1, underscoring the role of altitude in mitigating routing contention. Furthermore, for a 1 MB burst size, Telesat approaches the performance of Starlink, which holds more than seven times the number of satellites. Finally, Telesat reaches a burst loss reduction of 18.4\% by using a 1 MB burst size (2.04\% loss), compared to itself with 50 MB bursts (2.5\% loss), demonstrating that appropriate burst sizing can substantially reduce resource contention. In general, burst contention is the main source of burst losses based on the zoom-in in Fig.\ref{fig:blocking}.

\begin{figure}[ht]
    \centering
    \includegraphics[width=0.97 \linewidth]{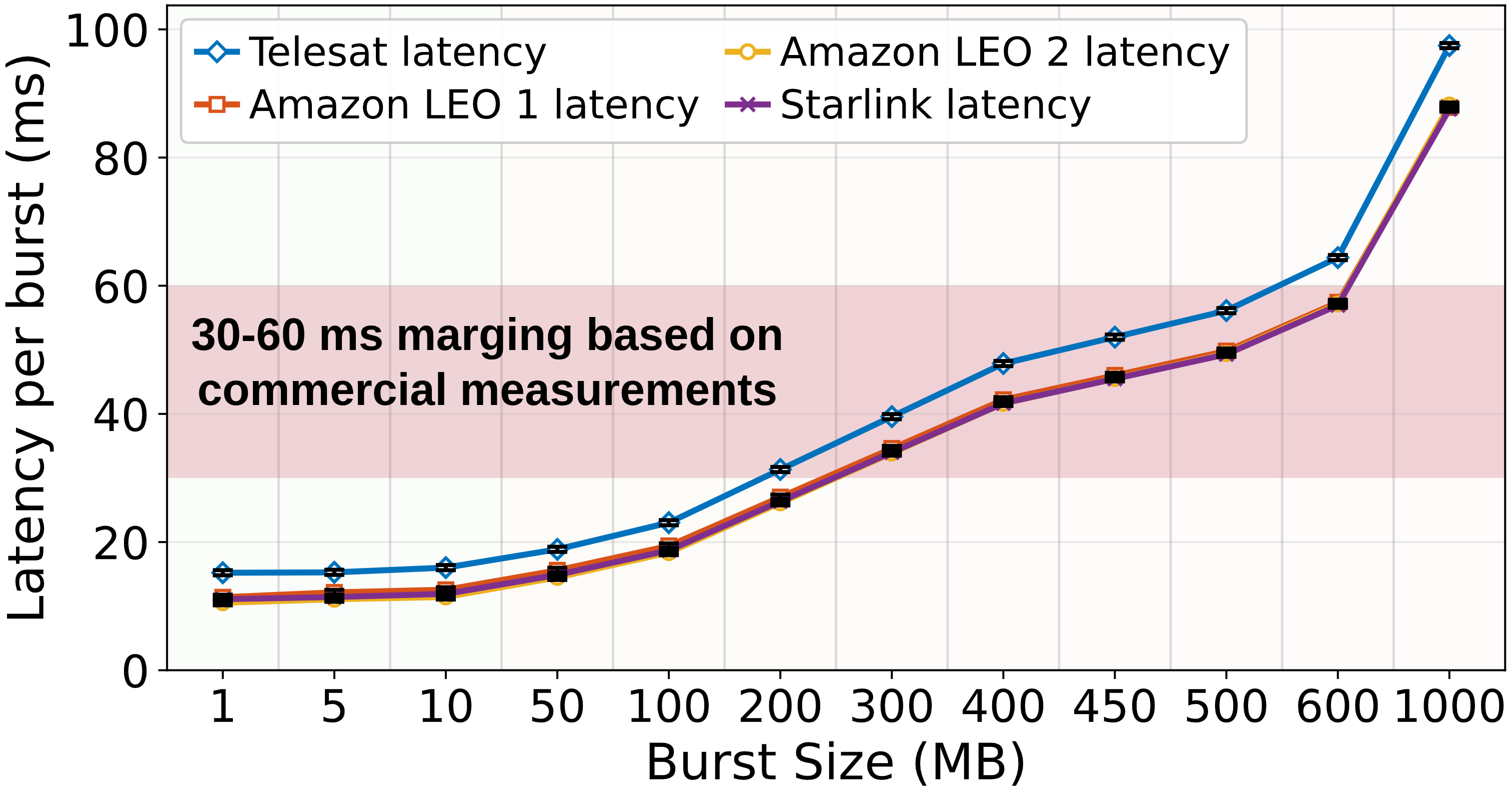}
    \caption{E2E latency for different burst sizes and constellations.}
    \label{fig:latency_curves}
\end{figure}

As shown in Fig.~\ref{fig:latency_curves}, E2E latency has a strong linear dependence on the burst size, and the lowest latency points occur at the smallest evaluated burst sizes across all constellations. The Telesat constellation,  due to its higher altitude, has an average 14.74\% higher E2E latency than other constellations. Fig.~\ref{fig:latency} further decomposes the E2E latency into its components, clarifying the relative contribution of each delay. 

\begin{figure}[ht]
    \centering
    \includegraphics[width=0.99 \linewidth]{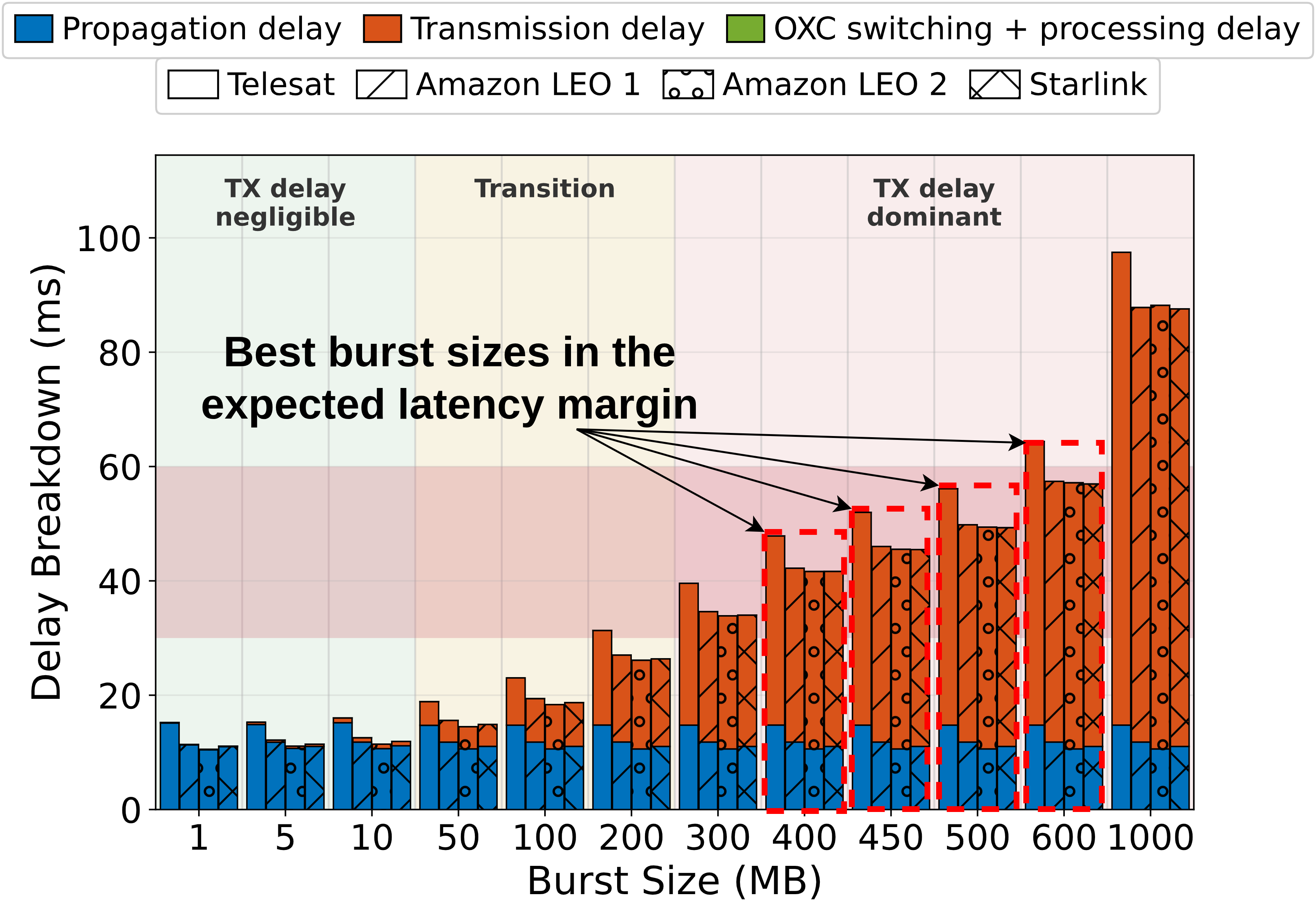}
    \caption{E2E latency breakdown for different burst sizes and constellations.}
    \label{fig:latency}
\end{figure}
Latency decomposition reveals that the dominant delay component arises from burst transmission for burst sizes larger than 200 MB. Specifically, transmission delay scales linearly with burst size, as larger data payloads require proportionally longer laser transmission times. Considering the 30 - 60 ms delay range experienced by modern satellite constellations, the latency results indicate that feasible burst sizes range from 300 to 600 MB across all lower altitude LEO constellations, and 200 to 500 MB for Telesat, as its increased altitude adds to propagation delay and a smaller burst size is required to reduce the transmission delay.

\section{Conclusions}
We evaluated the maximum feasible burst size in all-optical satellite networks using the OBS switching paradigm, under burst loss rate and latency constraints for different satellite topologies. Simulation results highlight that the maximal feasible burst size for the bottleneck of 100 Gbps is defined as 500 MB for higher altitude LEO constellations and 600 MB for lower altitude ones, while more dense constellations can also lead to a more efficient network by reducing the burst loss ratio.


\clearpage
\section{Acknowledgements}
This work was partly funded by the European Space Agency (ESA) in the framework of the project NEXON (activity number 1000042074).

\defbibnote{myprenote}{%

}
\printbibliography[prenote=myprenote]

\vspace{-4mm}

\end{document}